\documentclass{article}
\usepackage{spconf,amsmath,graphicx,amssymb}
\usepackage{color,multirow}
\usepackage{cite}
\usepackage{type1cm}
\usepackage{algorithm}
\usepackage{algorithmic}
\usepackage{setspace}   % for the draft mode
\DeclareMathOperator\SNR{SNR}

\DeclareMathOperator\DS{DS}
\newcommand{\mysubsection}[1]{\vspace{-8pt}\subsection{#1}}
\title{Source Mixing and Separation Robust Audio Steganography}
\name{Naoya Takahashi, Mayank Kumar Singh, Yuki Mitsufuji}
\address{Sony Group Corporation, Japan}
\begin{document}
\ninept
% \fontsize{9pt}{10pt}\selectfont
% \fontsize{9pt}{22pt}\selectfont   % for review

% \begin{onecolumn} % for the draft mode
% \begin{spacing}{3} % for the draft mode
% \large % for the draft mode 

\maketitle
\begin{abstract}
Audio steganography aims at concealing secret information in carrier audio with imperceptible modification on the carrier. Although previous works addressed the robustness of concealed message recovery against distortions introduced during transmission, they do not address the robustness against aggressive editing such as mixing of other audio sources and source separation. 
In this work, we propose for the first time a steganography method that can embed information into individual sound sources in a mixture such as instrumental tracks in music. To this end, we propose a time-domain model and curriculum learning essential to learn to decode the concealed message from the separated sources. Experimental results show that the proposed method successfully conceals the information in an imperceptible perturbation and that the information can be correctly recovered even after mixing of other sources and separation by a source separation algorithm. Furthermore, we show that the proposed method can be applied to multiple sources simultaneously without interfering with the decoder for other sources even after the sources are mixed and separated.
\end{abstract}
\section{Introduction}
\label{sec:intro}
Audio steganography\cite{KAR201554,Kreuk20HideSpeak,Wu2020stegano} is the science of concealing secret messages inside a host audio called a \textit{carrier} in such a way that the concealment is unnoticeable to human ears. Recently, deep neural networks (DNNs) have been used as a steganographic function for hiding data inside images to achieve high \textit{capacity} \cite{Baluja17,Hayes17,Zhu18HiDDeN,Lu21Stegano}. Kreuk et al. successfully adopted a DNN-based steganographic approach for audio by concealing a message in the short-time Fourier transform (STFT) domain while considering the distortion caused by the inverse STFT with a mismatched phase \cite{Kreuk20HideSpeak}. Although the method demonstrates robustness against some types of distortion that can be introduced during data transmission, the method does not assume that the carrier could be aggressively edited. Therefore, it is difficult to conceal messages in each source in a mixture, such as individual instrumental tracks in music, and recover the messages from the source-separated sounds of the mixture.  
In this work, we address this problem for the first time and propose a DNN-based steganographic method that works even after source mixing and separation.

% Thanks to the advances in source separation methods based on DNNs, the accuracy of source separation has been dramatically improved
% \cite{JanssonHMBKW17,Luo18cTAS, JLee2019,Liu2019mss,defossez2019music, Takahashi21}.
% \cite{JanssonHMBKW17,Takahashi18MMDenseLSTM,JLee2019,Liu2019mss,defossez2019music,Takahashi19, Takahashi21}.
Recently, as the source separation accuracy has been considerably improved owing to advances in DNN-based methods 
% \cite{JanssonHMBKW17,Luo18cTAS, JLee2019,Liu2019mss,defossez2019music, Takahashi21}, 
\cite{JanssonHMBKW17,Luo18cTAS, Takahashi18MMDenseLSTM, JLee2019,Liu2019mss,defossez2019music,Takahashi19,Takahashi21}, 
separated sources are being widely used for many purposes such as karaoke, creating new contents using separated sources, and training models using a dataset that consists of separated sounds \cite{Liu2020, Basak20}.
Consequently, it is becoming more important to focus on the value and functionality of sound sources, which may be mixed with other sources and later separated by a source separation algorithm.
The proposed method addresses this issue and can be used for various applications.
For instance, secret communication, the main focus of stenography, can be extended to source-wise communication where messages are concealed in sound sources and mixed with other unknown sources. Recipients who are aware of the presence of messages can decode them from individual separated sources. 
% The combination of source types and concealed messages can provide extra capacity. 
Sound source creators can conceal information such as musical notes of the source, captioning of the source, copyright information, or any unrelated information. In the case that copyright information is concealed for ownership protection, the method is also called watermarking. The protection of creators' rights against the abuse of separated sources is also becoming increasingly important and our proposed method addresses this problem.

In this work, we focus on musical sources. Our goal is to enable creators to conceal information in sound sources independently. Therefore, we mainly focus on the imperceptibility of the modification of the sources and the robustness against source mixing and separation.  
% Our research focus is to enable audio steganography Although the copyright protection of sound sources is one of the important usage, we greatly care about the imperceptibility as our primary target is professional music contents.
% Moreover, we mainly focus on the robustness against the distortion introduced by the separation rather than data transmission because we prioritize the protection of high fidelity, not degraded sources against the undesired secondary use in digital domain.
The contributions of this work are fivefold:
{
\setlength{\leftmargini}{12pt} 
\begin{enumerate}
\item We propose concealing messages inside a source that will be mixed with other \textit{unknown} sources and then separated by the source separation method. We call the proposed method source mixing and separation robust audio steganography (MSRAS).
\item We propose a DNN-based concealer to conceal messages in the time domain to avoid distortions caused by phase mismatch and enable simple end-to-end concealer and decoder optimization through the source separation model.
\item We further propose curriculum learning, which is shown to be essential to train the concealer and decoder through the source separation model.  
\item We empirically show that MSRAS can recover a message from both unprocessed and separated sources with high accuracy while the modifications for the concealment are hardly detectable by human ears. We also show the robustness of the proposed method against other types of noise.
\item We further show that MSRAS can be applied to multiple sources in a mixture to conceal messages independently without interfering with the messages in other sources. This enables creators to hide messages in each source independently without knowing other sources that they will be mixed with. 
\end{enumerate}
}

\section{Related works}
As mentioned above, steganography is closely related to watermarking. While the main goal of steganography is secret communication and it focuses on imperceptibility, watermarking focuses more on robustness and is typically used for ownership protection and verification.
% Audio watermarking aims at concealing data called \textit{messages} inside the host audio  for ownership protection and verification.
Various audio watermark approaches have been proposed, such as patchwork \cite{Kalantari09,Xiang14}, spread spectrum \cite{Valizadeh12,Liu03}, echo-hiding \cite{YXiang11,TCChen08}, support vector regression \cite{XWang07,Lakshmi08}, and singular value decomposition \cite{Dhar15}. Recently, deep-learning-based methods have been proposed for image watermarking and steganography \cite{Zhu18HiDDeN}.

Our work is also related to adversarial examples, where DNNs are shown to \textit{detect} imperceptible features that can be designed to manipulate network prediction \cite{Szegedy2014,Ilyas2019}. Recently, adversarial examples have been explored in audio source separation \cite{Takahashi21}.

% blind-water marking.

\section{Proposed method}
% \label{sec:amicable}
\mysubsection{Audio steganography}
The steganographic model consists of a concealer $E(c,m) = \hat{c}$, which conceals a message $m$ inside a carrier audio signal $c$, and a decoder $D$ which recovers the message from the embedded carrier as $D(\hat{c})=\hat{m}$.  The goal of the concealer and decoder is to minimize the message reconstruction error $d_m(m,\hat{m})$, while regularizing the modification of the carrier to be minimal. The model is trained by minimizing the loss function $L$ defined as

\begin{equation}
    L(c,m) =\lambda_c d_c(c, \hat{c})+\lambda_md_m(m,\hat{m}),
    \label{eq:stegano}
\end{equation}
where $\lambda_c$ and $\lambda_m$ denote the weights, and $d_c(c, \hat{c})$ is the metric used to measure the similarity between the original and embedded carriers. In \cite{Kreuk20HideSpeak}, the $l1$ norm is used for both $d_c$ and $d_m$.

\begin{figure}[t]
  \centering
  \includegraphics[width=\linewidth]{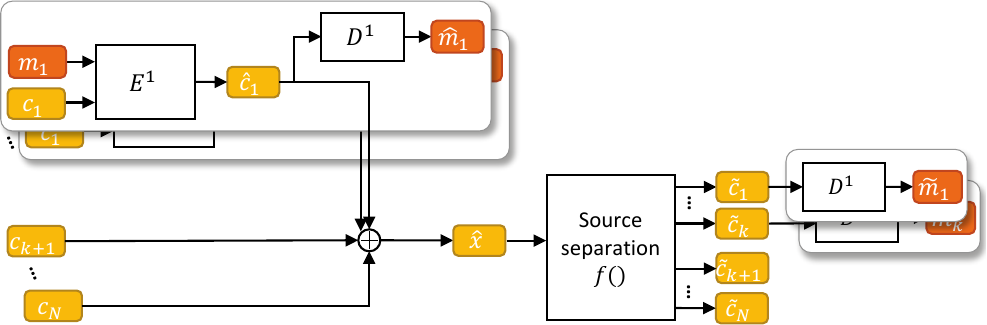}
  \caption{Model overview.}
  \label{fig:overview}
\end{figure}

\mysubsection{Incorporating source mixing and separation}
\label{sec:msras}
We extend the method described in the previous section to enable message concealment in individual sources of a mixture so that the message can be recovered from the separated sources, as shown in Fig.~\ref{fig:overview}. For simplicity, we assume that source separation is performed on the basis of the source type such as the instrument type in music. Given $N$ sources $c_1, \cdots, c_N$ of the mixture $x = \sum_{n=1}^N c_n$, we conceal messages $m_{i\in\Gamma}$ into $k$ of the $N$ sources. Without loss of generality, we choose the first $k$ sources as the ones containing messages ($\Gamma=[1,\cdots,k]$). The concealer $E^i$ takes the source $c_i$ and message $m_i$ as inputs and it outputs the embedded source $\hat{c}_i=E^i(c_i, m_i)$ independently from other sources. The embedded and non-embedded sources are mixed to form the mixture $\hat{x}$, which can be a final product to be distributed (such as music). Source separation $f(\hat{x})$ is then applied to obtain the separation $\Tilde{c}_i$. Our goal is to recover the messages from both $\hat{c}_i$ and $\Tilde{c}_i$ using the decoder $D^i()$. The loss function for source $i$ becomes
\begin{equation}
    L^{(i)} =\lambda_1^i d_c(c_i, \hat{c}_i)+\lambda_2^id_m(m_i,\hat{m}_i)+\lambda_3^id_m(m_i,\Tilde{m}_i),
    \label{eq:msrar}
\end{equation}
where $\Tilde{m}_i = D^i(f(\hat{x})_i)$ denotes the message recovered from the $i$th separated source $f(\hat{x})_i$.
% where $\Tilde{m}_i = D^i(f(\hat{x})_i)$ and $\hat{x}=\sum_{j\in\Gamma}\hat{c}_i+\sum_{j\in\Bar{\Gamma}}c_j$.
Note that since $\Tilde{m}$ depends on all sources in the mixture $\hat{x}$, the loss $L^{(i)}$ is also related to other concealers $E^{j\neq i}$. One way to train all encoders and decoders is to use the summation of all losses $L^{all}=\sum_{i\in\Gamma}L^{(i)}$ universally. However, this strategy unnecessarily promotes the dependence on other unobserved sources for each concealer and failed to learn any steganographic function. Therefore, we instead use only $L^{(i)}$ to train the model for source $i$ by freezing other concealers and alternately train the models for each source as shown in Algorithm \ref{alg1}. 
% This training scheme promotes the independence of each model from other sources while being aware of other concealers' concealment \textit{strategies} to avoid interference, which enables both the message sender and the receiver to work on the source without knowing the other sources. 
This training scheme promotes the independence of each model from other sources and avoids interference by knowing the concealment \textit{strategies} of other concealers. This allows both the sender and the receiver of the message to work on the source without knowing the other sources.
\begin{algorithm}                      
    \caption{ Multi-source concealer decoder learning}         
    \label{alg1}
    \footnotesize
    \begin{algorithmic}[1]
    \FOR{ (\# iteration) } 
        \STATE sample $c_1,\cdots,c_N$ and  $m_1,\cdots,m_k$
        \FOR{ $i=1,...,k$ (source types) } 
            \STATE // conceal messages and mix 
            \STATE $\hat{x}=\sum_{j=1}^kE^j(c_, m_j)+\sum_{j=k+1}^Nc_j$
            \STATE // decode messages from $i$th embedded source and separation 
            \STATE $\hat{m}_i = D^i(\hat{c}), \Tilde{m}_i = D^i(f(\hat{x})_i)$
            \STATE freeze model parameters of $E^{j\neq i}, D^{j\neq i}$ and update the parameters of $E^{i}$ and $D^{i}$ using $L^{(i)}$ in \eqref{eq:msrar}
        \ENDFOR
    \ENDFOR  
    \end{algorithmic}
\end{algorithm}

% Note that both concealers and decoders are independent from that of other sources, which enables both message sender and receiver to work on the source without knowing the other sources. 

\begin{figure}[t]
  \centering
  \includegraphics[width=\linewidth]{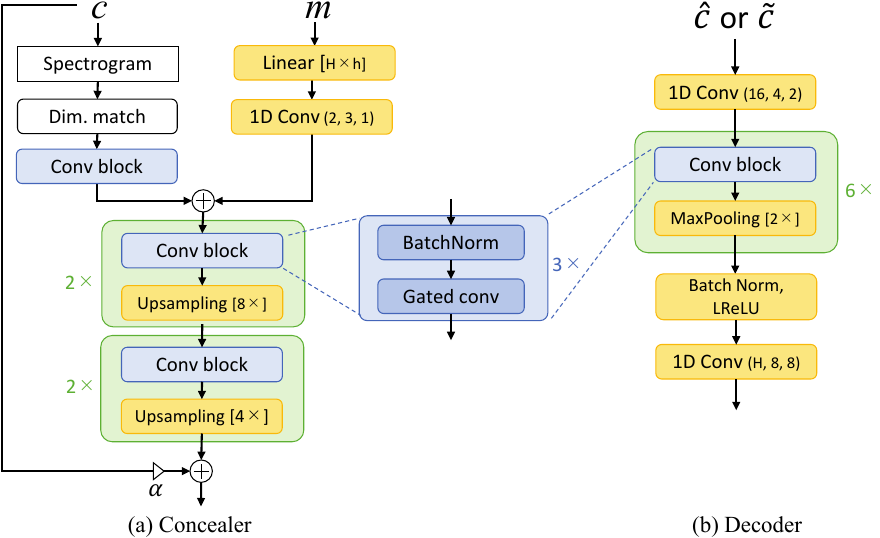}
  \caption{TCD model architecture.  Conv ($o,\kappa,s$) denotes the convolution with $o$ filters, kernel size $\kappa$, and stride $s$.}
  \label{fig:archi}
\end{figure}

\mysubsection{Concealing message in time domain}
Kreuk et al.~conceal and decode the message in the spectrogram domain \cite{Kreuk20HideSpeak}, where the embedded spectrogram $\hat{C}$ and the phase of the original carrier $\angle C$ are used to apply the inverse STFT to recover the time-domain signal. The mismatch of the magnitude and phase produces distortion in the embedded spectrogram $\hat{C}$. This problem is addressed by using the distorted spectrogram during the decoder training to model the distortion.

Recently, generative adversarial networks have been shown to generate a high-quality waveform from a mel-spectrogram \cite{Kumar19MelGAN}. Motivated by this work, we propose to directly conceal the message in the time domain.
To this end, we propose a network architecture for a time domain concealer as shown in Fig.~\ref{fig:archi}. The concealer projects the message to $h$-dimensional embeddings, and a 1D convolution is applied to it. 
The carrier (possibly multi-channel) signal is first converted to a spectrogram and the time and frequency dimensions are downsampled/interpolated to match the dimensions of the message embeddings. After a \textit{conv block}, which comprises three stacks of Batch Norm and gated convolution \cite{Dauphin17}, the output is merged with the message embeddings. Then, the conv block and upsampling layers are alternately applied four times to match the time dimension of the input waveform. The upsampling layers consist of the transposed convolution and we halve the number of channels at each layer. The output of the final layer is then mixed with the input carrier weighted by the mixing ratio $\alpha$.  
The decoder consists of a 1D convolution, followed by six stacks of conv blocks and max pooling layers, a Batch Norm, leaky ReLU non-linearity, and finally another 1D convolution. We refer to the proposed architecture as the time-domain concealer and decoder (TCD).
In our experiment, we consider byte data or, more specifically, a sequence of $H$-dimensional one-hot-vectors, with a frame rate of $T$ as the message. Therefore, the output of the decoder is fed to a softmax to obtain the posterior over the $H$ classes. However, the method can be directly applied to other message data such as an $H$-band mel-spectrogram with a frame rate of $T$.

\begin{table}[t]
    \caption{\label{tab:curriculum} {\it Curriculum for model training.}}
    \vspace{2mm}
    \centering{
      \footnotesize
    \begin{tabular}{c | c c c c c c} 
    \hline
    Step & $\lambda_1$	& $\lambda_2$	& $\lambda_3$	& $\alpha$	&  $d_c$ & \# iteration\\
    \hline\hline
    1 &	1   & 0 & 0 & 0 & MDS   & 500\\
    2 &	0.1   & 1 & 0 & 0 & MDS   & 2500\\
    3 &	0.1   & 1 & 1 & 0 & MDS   & 5000\\
    4 &	0.1   & 1 & 1 & 0$\rightarrow$1 & MDS   & 2000\\
    5 &	0.0002   & 1 & 1 & 1 & MDSR   & 5000\\

    \hline
    \end{tabular}
    }
\end{table}

\mysubsection{Curriculum learning}
Since source separation is a highly nonlinear process due to it being an ill-posed problem, learning the TCD through the source separation model is extremely challenging. The models are unable to learn the steganographic function from scratch by directly using \eqref{eq:msrar}. To mitigate this difficulty, we propose curriculum learning as follows:  \\
\textbf{Step 1:} \hspace{1mm} We begin by training the concealer to generate audio that sounds similar to the carrier because the embedding signal must pass through the source separation model to enable message recovery from the separated source. To this end, we introduce an auxiliary loss function called a multi-resolution downsampled spectral (MDS) loss and use it as the metric $d_c$: 
\begin{equation}
    MDS(c,\hat{c}) = \sum_{m\in M}\|\DS(\mathcal{S}^m(c)) - \DS(\mathcal{S}^m(\hat{c})) \|^2,
    \label{eq:mds}
\end{equation}
where $\DS$ denotes the average pooling function with a kernel size of $(32,1)$, $\mathcal{S}^m$ is a function used to compute a spectrogram with STFT parameter $m$, and $M=[1024, 2048, 4096, 8192]$ is the set of STFT window lengths. The idea of MDS is to promote a natural sound by employing a multi-resolution STFT loss \cite{Yamamoto20} while allowing local spectral differences by downsampling the spectrogram. We set the mixing ratio $\alpha$ in the concealer to zero to avoid the trivial solution.\\
\textbf{Step 2:} Once the concealer starts to generate sounds similar to the source, we start to conceal the message. However, we first focus on the message recovery from $\hat{c}$ only.\\
\textbf{Step 3:} \hspace{1mm}After the decoder learns how to recover the message from the embedded source $\hat{c}$, we introduce source separation to the training criteria.\\
\textbf{Step 4:} \hspace{1mm}We gradually increase $\alpha$ in Fig.~\ref{fig:archi} to one. This helps to minimize the perturbation from the original carrier.\\
\textbf{Step 5:} \hspace{1mm}To further ensure the imperceptibility of the perturbation, we switch $d_c()$ from MDS to the multi-resolution downsampled spectrogram ratio (MDSR) defined as  
\begin{equation}
    MDSR(c,\hat{c}) = \sum_{m\in M}\|\DS(\mathcal{S}^m(c-\hat{c})) / \DS(\mathcal{S}^m(c)) \|.
    \label{eq:mdsr}
\end{equation}
The MDSR promotes the ratio of the downsampled spectrogram of the carrier and the perturbation to be small. Thus, when the carrier contains high energy in a time-frequency band, the perturbation can also have higher energy in that band, which approximately considers the masking effect. The curriculum is summarized in Table \ref{tab:curriculum}.

\mysubsection{Increasing robustness}
\label{sec:aug}
% Although our main focus is the robustness against the source mixing and separation, we also consider to improve the robustness against other types of distortion. 
% To this end, 
To improve the robustness against noise, we use three types of data augmentation during step 5 to improve the robustness of the model. (i) \textit{Channel masking}: We randomly mask half the length of the carrier signal from one of the channels (we use stereo audio in our experiment). (ii) \textit{Additive Noise}: We add Gaussian noise with $\sigma=0.001$. (iii) \textit{Random EQ}: A low-pass filter with a random cutoff frequency (15$\sim$20 kHz) is applied to half of the samples in the batch.

\section{Experiments}
\mysubsection{Setup}
Experiments are conducted on the MUSDB18 dataset \cite{sisec2018}, which contains 100 and 50 songs for the \textit{train} and \textit{test} sets, respectively.  
For each song, four sources ({\it bass, drums, other, vocals}) and their mixture recorded in stereo format at 44.1 kHz are available. Unless otherwise noted, we use the \textit{drum} track as the carrier to conceal the message.
We consider text data as the message. We encode the 26 letters of the alphabets and one end-token to 27-dimensional one-hot-vectors every 23.2 ms; thus, the message capacity is about 43 characters per second. % or 205 bps.
Randomly sampled character sequences are used for both training and testing.
As the source separator, we use Demucs \cite{defossez2019music}, which is an open-source music source separation library and performs separation in the time domain. We use the provided pre-trained weights without any modification.
As the metric of message $d_m$, we use the cross-entropy loss.
Models are trained on the \textit{train} set using the Adam optimizer with a learning rate of 0.001 and a batch size of 12, and are evaluated on the \textit{test} set.

\begin{table}[t]
    \caption{\label{tab:obj} {\it SNR and accuracy of different methods.}}
    \vspace{2mm}
    \centering{
      \footnotesize
    \begin{tabular}{c | c c | c c} 
    \hline
    \multirow{2}{*}{Model} & \multicolumn{2}{c|}{SNR [dB]} & \multicolumn{2}{c}{Accuracy [\%]}\\
        % & Embedded $\hat{c}$  & Separated $\Tilde{c}$ & Embedded $\hat{c}$  & Separated $\Tilde{c}$\\
        & $\hat{c}$  & $\Tilde{c}$ & D($\hat{c})$  & D($\Tilde{c}$)\\
    \hline\hline
    % Random guess & - & - & 3.7  & 3.7\\
    Hide\&Speak\cite{Kreuk20HideSpeak} &  30.5 &   25.9  &  \textbf{100.0} &  36.0\\
    TCD trained with \eqref{eq:stegano} & 35.3 & 40.3 & \textbf{100.0} & 3.6\\
    Hide\&Speak$+d_m(m_i,\Tilde{m}_i)$ & 31.6 &   25.3    & 99.6  &   74.7\\
    MSRAS ($\lambda_2=0$) & 35.3 & \textbf{37.8}  & 61.5  & \textbf{100.0}\\
    \hline
    MSRAS (Proposed) & \textbf{35.9} & 34.5  &  \textbf{100.0} & \textbf{100.0}\\
    \hline
    \end{tabular}
    }
\end{table}

\mysubsection{Objective evaluation}
We evaluate the proposed method by comparing with baselines. 
% Unfortunately, only limited works are available on audio steganography. 
We adopt the state-of-the-art method in \cite{Kreuk20HideSpeak}, which was originally proposed for hiding a spectrogram in another spectrogram, by adding a linear layer at the beginning of the concealer and the end of the decoder to match the dimension of the message to that of the spectrogram. We also consider three other baselines: (i) the method in \cite{Kreuk20HideSpeak} is extended to incorporate the loss on the message recovered from the separation expressed as \eqref{eq:msrar},  (ii) the proposed time-domain model, the TCD, is trained using the conventional loss \eqref{eq:stegano}, (iii) the proposed model is trained only on the separated source ($\lambda_2=0$). To evaluate the distortion of sources, we report the signal-to-noise ratio $\SNR$(\textit{signal}, \textit{noise}) on both the embedded source and separated source as $\SNR$($c$, $c-\hat{c}$) and $\SNR$($f(c)$, $f(c)-\Tilde{c}$), respectively. 
Generally, SNR and accuracy are in the relationship of trade-off. We tune weight parameters $\lambda$ such that SNR of the embedded source becomes higher than 30~dB as the perturbation becomes hardly audible around that level.
The results are shown in Table \ref{tab:obj}. 
Both Hide\&Speak\cite{Kreuk20HideSpeak} and the TCD achieve high accuracy with a high SNR on the embedded source $\hat{c}$. However, they fail to decode the message after it is mixed with other sources and separated ($\Tilde{c}$). By incorporating the loss on the recovered message from the separation ($d_m(m_i,\Tilde{m}_i)$) into Hide\&Speak\cite{Kreuk20HideSpeak}, the decoding accuracy from the separated source is improved, but it remains significantly lower than the accuracy on $D(\hat{c})$. The proposed method achieves 100\% accuracy on both the embedded and separated sources with high SNRs. This highlights the robustness of the TCD against source mixing and separation. Interestingly, when we train the TCD to recover the concealed message only from the separated source ($\lambda_2=0$), we obtain 61.5\% accuracy on the embedded source even though 100\% accuracy is obtained on the separated source. We also train Hide\&Speak in the same setting of $\lambda_2=0$; however, it fails to learn any steganographic functions and the accuracy is the same as that of random guessing (1/27=3.7\%). These results further validate the effectiveness of the proposed TCD. 

Since both the proposed method and Hide\&Speak baseline obtain 100\% accuracy on $\hat{c}$, the difference is not visible. We further compare these models on a harder task by increasing the type of letters to 96. With the SNR on $\hat{c}$ being around 36 dB, Hide\&Speak and the proposed method obtain 99.3\% and 98.8\% accuracy on $D(\hat{c})$, and 8.1\% and 97.7\% on $D(\Tilde{c})$, respectively. This results show that the proposed method performs competitively on the embedded sources, yet performs much more robustly on the separated sources.

\mysubsection{Subjective detectability test}
We also conduct a subjective test to evaluate the perceptibility of the perturbation. An ABX test is performed for both the embedded source $\hat{c}$ and separated source $\Tilde{c}$, where in case of $\hat{c}$, X is the original source $c$ and either A or B is the same as X and the other is $\hat{c}$, whereas in case of $\Tilde{c}$, X is the separation of the mixture of the original sources. Subjects are asked to identify which one of A or B is the same as X, and allowed to listen to the samples many times. Forty-four audio engineers evaluate 10 samples of three second audio for each case, resulting in 440 evaluations. 
As shown in Table \ref{tab:subj}, the accuracy of correctly identifying the unmodified source is close to the chance rate (50\%); thus, we conclude that the distortion caused by the message concealment is hardly perceptible for human ears.

\begin{table}[t]
    \caption{\label{tab:subj} {\it Subjective test on perceptibility of the perturbations.}}
    \vspace{2mm}
    \centering{
    \footnotesize
    \begin{tabular}{c | c } 
    \hline
    Evaluated source & Accuracy [\%]\\
    \hline\hline
    Embedded source ($\hat{c}$) & 55.5\\
    Separated source ($\Tilde{c}$) & 53.9\\
    \hline
    \end{tabular}
    }
\end{table}

\begin{table}[t]
    \caption{\label{tab:excurriculum} {\it Comparison of different curricula.}}
    \vspace{2mm}
    \centering{
      \footnotesize
    \begin{tabular}{c | c c | c c} 
    \hline
    \multirow{2}{*}{Step} & \multicolumn{2}{c|}{SNR [dB]} & \multicolumn{2}{c}{Accuracy [\%]}\\
        % & Embedded $\hat{c}$  & Separated $\Tilde{c}$ & Embedded $\hat{c}$  & Separated $\Tilde{c}$\\
        & $\hat{c}$  & $\Tilde{c}$ & D($\hat{c})$  & D($\Tilde{c}$)\\
    \hline\hline
    2,3,4,5 & 54.1 & 52.1 & 6.3  & 6.3\\
    1,3,4,5 & 74.3 & 70.9 & 6.3  & 6.3\\
    1,2,4,5 & 27.2 & 26.3 & 100.0 & 100.0\\
    1,2,3,5 & 27.4 & 28.2  & 100.0 & 100.0\\
    1,2,3,4 & 18.3 & 20.9  & 100.0  & 100.0\\
    All & 35.9 & 34.5  &  100.0 & 100.0\\
    \hline
    \end{tabular}
    }
\end{table}

\mysubsection{Curriculum learning}
Next, we show the effectiveness of the proposed curriculum learning. To this end, we omit one of the steps in the curriculum and compare the SNR and accuracy. When we omit the step, we extend the number of iterations at the next step to match the total number of iterations for fair comparison. As shown in Table \ref{tab:excurriculum}, steps 1 and 2 are essential for the TDC to learn the steganographic function. Omitting step 3, 4, or 5 does not deteriorate the accuracy, however, SNR decreases. Therefore, we conclude that steps 3 to 5 help to learn a highly imperceptible concealment function.

\mysubsection{Interference from multiple models}
We also test the case where multiple sources contain embedded messages. We train models for \textit{drums} and \textit{vocals} tracks as described in Sec. \ref{sec:msras}. The results are shown in Table \ref{tab:multi}. The accuracy of message recovery from \textit{drums} separation remains high even if we include the concealer for \textit{vocals}. Although the accuracy of \textit{vocals} separation is relatively low compared with that of \textit{drums}, the accuracy is similar to that on the embedded source (D($\hat{c}$)$\approx$D($\Tilde{c}$)). This shows that the messages can be concealed in multiple sources and recovered from each separated source without interfering with the other models.

\begin{table}[t]
    \caption{\label{tab:multi} {\it Multiple models.}}
    \vspace{2mm}
    \centering{
    \footnotesize
\begin{tabular}{c | c c | c c} 
    \hline
    \multirow{2}{*}{Instruments} & \multicolumn{2}{c|}{SNR [dB]} & \multicolumn{2}{c}{Accuracy [\%]}\\
        % & Embedded $\hat{c}$  & Separated $\Tilde{c}$ & Embedded $\hat{c}$  & Separated $\Tilde{c}$\\
        & $\hat{c}$  & $\Tilde{c}$ & D($\hat{c})$  & D($\Tilde{c}$)\\
    \hline\hline
    drums & 27.9    & 30.8 & 100.0  & 100.0\\
    vocals & 32.3	& 29.9	& 86.8 & 85.6\\
    \hline
    \end{tabular}
    }
\end{table}

\begin{table}[t]
    \caption{\label{tab:robust} {\it Robustness of models trained with and without data augmentation evaluated with different types of distortion.}}
    \vspace{2mm}
    \centering{
    \footnotesize
    \begin{tabular}{c | c c } 
    \hline
    Distortion & w/o D.A. & w/ D.A.\\
    \hline\hline
    % None &  \textbf{100.0}   & 97.7 \\
    AGN ($\sigma=0.001$) &  \textbf{99.9}   & 97.2 \\
    Ch. drop  &  3.4  & \textbf{97.0} \\    
    EQ  &  \textbf{100.0}  & 97.1 \\    
    MP3 compression 320k &  3.4   & \textbf{95.0} \\    
    MP3 compression 128k &  3.7   & \textbf{94.9} \\    
    \hline
    \end{tabular}
    }
\end{table}

\mysubsection{Robustness against noise and edits}
Finally, we investigate the robustness against different types of perturbation, namely, additive Gaussian noise (AGN), a channel drop that masks one of the channels, five-band-equalization with random gain from $-$3 to 3 dB (EQ), and MP3 compression. The noises are applied to the mixture $\hat{x}$, and the separation of the distorted mixture is tested. The accuracy of two models trained with and without the data augmentation described in Sec. \ref{sec:aug} is shown in Table \ref{tab:robust}. The model trained without the data augmentation is robust against additive noise and equalization but vulnerable to data loss and signal compression. The data augmentation is shown to greatly improve the robustness against these types of noise and edits.

\section{Conclusion}
We propose audio steganography that is robust against source mixing and separation, where messages are concealed in some sources individually and the messages are recovered after mixing with other sources and separation of the mixture. To this end, we propose curriculum learning for training a time domain concealer and decoder. Experimental results confirm the effectiveness of the proposed method from different perspectives. Future works include generalizing the method to unseen separation models, testing on other domain signals such as a mixture of speeches and environmental sounds, and exploring the ability to evade steganalysis methods.
% other types of editing and unseen separation algorithms.

\ninept

% -------------------------------------------------------------------------
\bibliographystyle{IEEEbib}
\bibliography{watermark,MIR,bss,other,advexamp}

% \end{spacing} %for the draft mode
% \end{onecolumn} %for the draft mode

\end{document}